\documentclass[aps,showpacs,12pt,nofootinbib]{revtex4}

\usepackage{epsfig}
\usepackage{graphicx,graphics}
\usepackage{enumerate}
\usepackage{subfigure}
\usepackage{amsmath}
\usepackage{amsfonts}
\usepackage{amssymb}
\usepackage{amsthm}
\usepackage{psfrag}
\usepackage{dsfont}
\usepackage{color}
\def\G{\mathcal{G}}
\def\A{\mathcal{A}}
\def\ra{\rightarrow}

\newcommand\equal{&\!\!\!=\!\!\!&}

\def\d{\rm d}

\begin{document}

\title{Three-leg correlations in the two component spanning tree \\ on the upper half-plane}

\author{S.Y. Grigorev$^1$, V.S. Poghosyan$^{1,2}$, V.B. Priezzhev$^1$}
\affiliation{
$^1$Bogoliubov Laboratory of Theoretical Physics, JINR, 141980 Dubna, Russia\\
$^2$ Institute of Applied Problems in Physics, 25 Hr. Nersessian
Str., 375014 Yerevan, Armenia }
\begin{abstract}
We present a detailed asymptotic analysis of correlation functions
for the two component spanning tree on the two-dimensional lattice when one component contains three paths connecting  vicinities of two fixed lattice sites at large distance $s$ apart. We extend the known result for correlations on the plane to the case of the upper half-plane with
closed and open boundary conditions. We found asymptotics of correlations
for distance $r$ from the boundary to one of the fixed lattice sites for the cases $r\gg s \gg 1$ and $s \gg r \gg 1$.
\end{abstract}

\maketitle

\section{Introduction}
In recent years the logarithmic conformal field theories (LCFT) \cite{cft},\cite{lcft} and their relation to lattice models of statistical physics like dense polymers
\cite{SaleurN,polym,log}, the sandpile model \cite{sand,diss,bound,jpr}, dimer
models \cite{dim} and percolation \cite{perc,log} have been the subject of active research.
 Among all these models, the Abelian sandpile model
(ASM) \cite{dhar-prl} is one of the most interesting and fruitful, because
correlation functions containing logarithmic corrections can be found explicitly using combinatorial
methods. In this way, the full correspondence between the lattice model and the logarithmic conformal field theory can be transparently tested. A lot of successful checks have been made in \cite{sand,diss,bound,jpr}, including various calculations for correlations in the bulk and on the boundaries, the determination of boundary changing fields, the insertion of isolated dissipation and evaluation of some finite-size effects. One of the most popular questions and checks was about the origin of the logarithmic corrections to the correlations. It is known just two cause of that: the insertion of the dissipation at isolated sites \cite{diss}, that follows from the logarithmic behavior of the inverse Laplacian at a large distance, and non-locality of height variables for $h\geq 2$ in the ASM \cite{jpr}. The last case is more important and difficult, because sandpile configurations with height variables $h\geq 2$ are mapped onto an infinite set of non-local configurations of spanning trees. This non-locality arises due to the presence of a specific three-leg subgraph, so-called $\Theta$-graph \cite{Priez}. The $\Theta$-graph is a subconfiguration of the spanning tree, consisting of three paths, that connect the vicinity of vertex $j_0$ with that of vertex $t_0$ (Fig.\ref{ugolugol}). The paths with additional branches attached to them form one component, which is surrounded by another component of the spanning tree. A generalization of $\Theta$-graph is an odd ``$k$-leg'' subgraph, which has been considered by E.V.Ivashkevich and C.-K.Hu in \cite{IvaEugene} for the infinite square lattice. They obtained the asymptotic dependence  $P(r)\sim \ln r / r^{\frac{k^2-1}{2}}$, for $k=1,3,5,\ldots$ and concluded that it is the presence of the second component is responsible for the logarithmic correction to the correlation function. Indeed, H.Saleur and B.Duplantier considered correlations of ``k-leg operators'' by mapping of the two-dimensional percolation problem on a Coulomb gas and found, that the asymptotics of correlation functions for one component spanning tree has a pure power-law decay $r^{-\frac{k^2-1}{2}}$ \cite{Saleur}.

 Later on, G. Piroux and P. Ruelle calculated the height probabilities for $h\geq 2$ in the ASM on the upper-half plane with closed and open boundaries \cite{jpr}. They enumerated the spanning tree configurations with $\Theta$-graph, having one fixed site at distance $r$ apart the boundary and another site running over the whole upper half-plane. The summation over positions of the running site leads to cumbersome estimations of integrals, so that it is difficult to follow details of correlations between different parts of $\Theta$-graph explicitly. Calculations of two point correlations $P_{1h}-P_{1}P_h$ in the ASM on the plane for $h\geq2$ also lead to the same difficulties. The $\Theta$-graph arising for $h\geq 2$ consists of three paths connecting one fixed site with height $h\geq2$ with another site at the distance $s$,  running over the whole plane except the site with the height variable ``one''. It was shown in \cite{lette}, that evaluation of the logarithmic corrections to the two-point correlations does not need the summation by $s$ over the whole plane. Instead, it is enough to take into account only those $\Theta$-graphs configurations when the running site of the $\Theta$-graph is situated in a vicinity of the height variable ``one''. The latter approach, being much simpler, is not so transparent, and an additional analysis of correlations of different parts of the $\Theta$-graph is desirable. In this work, we find the asymptotic behavior of three-leg correlations for the case of the upper-half plane. We test validity of the method described in \cite{lette} for the upper half-plane, examining the order of expansion, where we can obtain a disagreement.


  \section{The model}
\begin{figure}[h!]
\includegraphics[width=200pt]{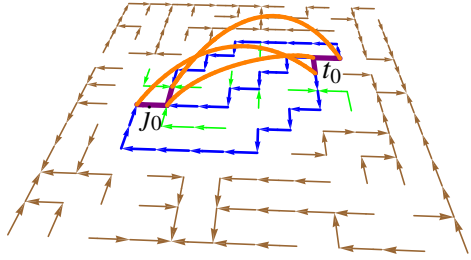}
 \caption{\label{ugolugol} Three-leg subgraph: the simplified version of $\Theta$-graph.}
\end{figure}

We consider the labeled graph $\mathcal{G} =(V,E)$ with vertex set
$V$ and set of bonds $E$. The vertices are sites of the square
lattice and an additional point which is the root ``$\star$'': $V\equiv \{s_{x,y}, (x,y) \in \mathbb{Z}^2, |x|\leq M, |y|\leq N \}\cup\{\star\}$. The bonds of $E$ connect
only neighboring sites. Vertices $i_{x_1,y_1}$, $j_{x_2,y_2} \in V $ are neighbors, if $(x_2-x_1)^2+(y_2-y_1)^2=1$. Also all boundary vertices $\{s_{x,\pm N}, x \in \mathbb{Z}\}$ and $\{s_{\pm M, y}, y \in \mathbb{Z}\}$ are neighbors of the root $\star$. The graph $\G$ represents a finite square
lattice of $2N+1 \times 2M+1$. In thermodynamical limit, $N,M\rightarrow \infty$, the lattice $s_{x,y}$
covers the whole two dimensional plane. We consider also the upper-half plane with closed and open boundary conditions at the
lattice sites $V_{0}=\{s_{x,1}, x \in \mathbb{Z}, |x|\leq M\}$. The corresponding graphs
are $\mathcal{G}_{cl}=(V_{+},E_{cl})$ and
$\mathcal{G}_{op}=(V_{+},E_{op})$, where $V_{+}=
\{s_{x,y}, x \in \mathbb{Z}, y \in \mathbb{N}, |x|\leq M, y\leq N\}\cup \{\star\}$, with the
 sets of bonds $E_{cl}\equiv E\cap \{(i,j),\ i\in V_{+},\ j\in V_{+}\}$ and $E_{op}\equiv E_{cl} \cup \{ (i,\star),\ i\in V_{0} \}$. We construct the desired
spanning tree configurations on the above graphs by using the arrow
representation, see e.g. \cite{Pr85}. Accordingly, we attach to each vertex $i
\in V\backslash \{\star\}$ an arrow directed along one of bonds $(i,i')\in E$
incident to it. Each arrow defines a directed bond $(i\ra i')$ and
each configuration of arrows $\A$ on $\G$ defines a spanning
directed graph (digraph) $\G_{dir}(\A)$  with set of bonds
$E_{dir}(\A)=\{(i\ra i'): i\in V\backslash\{\star\},\ i'\in V,\ (i,i')\in E\}$ depending on $\A$. Similarly,
the arrow configurations on $\G_{cl}$ and $\G_{op}$ define a spanning digraph
$\G_{cl,dir}(\A)$, $\G_{op,dir}(\A)$ with corresponding sets of bonds. Note that no arrow is attached to the root
$\star$, so that it has out-degree zero.
 A sequence of directed bonds $(i_1,i_2),(i_2,i_3),(i_3,i_4), \dots,
(i_{m-1},i_m)$ is called the path of length $m$ from the site $i_1$ to the site $i_m$. This path forms a loop if $i_m=i_1$. Spanning tree is a spanning digraph without any loops. Our aim is to construct a two-component spanning tree, with one component containing three paths connecting neighboring sites
$j_0,j_0-\hat{x},j_0+\hat{y}$ with sites $t_0,t_0-\hat{y},t_0+\hat{x}$, where
$j_0$ and $t_0$ have coordinates $(0,r)$ and $(k,l+r)$ respectively, and
$\hat{x}=(1,0)$, $\hat{y}=(0,1)$ are unit vectors
(Fig.\ref{ugolugol}). The relevant configurations will be investigated with aid of the determinant expansion of the Laplace matrix. This technique is described in details in \cite{jpr, dhar-prl, Priez}.
Let the vertices of the set $V$, be labeled in arbitrary order from $1$ to $n=|V\backslash\{\star\}|=(2M+1)(2N+1)$. Then Laplacian $\Delta$ of size $n \times n$ has the elements:
\begin{equation}
\label{Delta}
\Delta_{ij} =
\begin{cases}
  z_i & \text{if $i=j$},\\
 -1 & \text{if $i,j$ are nearest neighbors}, \\
  0 & \text{otherwise}.
\end{cases}
\end{equation}
where $z_i$ is the degree of vertex $i \in V\backslash \{\star\}$. The determinant of Laplace matrix is equal to the number of spanning trees on $\G$ with the root $\star$. Laplace matrices $\Delta_{op}$, $\Delta_{cl}$ for the upper half-plane have size $n=|V_+\backslash\{\star\}|=(2M+1)N$ and are defined by the same way as $\Delta$, but for graphs $\G_{op}$, $\G_{cl}$ respectively. The determinant of $\Delta$ is a sum over all permutations $\sigma$ of the set $\{1,2,\ldots,n\}$:
\begin{eqnarray}
\label{Leibniz}
\det\Delta=\sum_{\sigma \in S_n} \mathrm{sgn}(\sigma)\Delta_{1,\sigma(1)}\Delta_{2,\sigma(2)}\ldots\Delta_{n,\sigma(n)}
\end{eqnarray}
where $S_n$ is the symmetric group, $\mathrm{sgn}=\pm 1$ is the signature of permutation $\sigma$. In general, each permutation $\sigma \in S_n$ can be factorized into a composition of disjoint cyclic permutations, say, $\sigma=c_1\circ c_2\circ \ldots c_k$. This representation partitions the set of vertices $V\backslash\{\star\}$ into non-empty disjoint subsets which are orbits $\mathcal{O}=\{v_{i,1},v_{i,2}\ldots,v_{i,l_i}\}\subset V$ of the corresponding cycles $c_i$, $i=1,\ldots,k$, at that $\cup_{i=1}^k\mathcal{O}_i=V\backslash\{\star\}$ and $\sum_{i=1}^k l_i=n$, where $l_i$ is the length of cycle $c_i$. The orbits consisting of just one element, if any, constitute the set $S_{fp}(\sigma)$ of fixed points of the permutation: $S_{fp}(\sigma)=\{v=\sigma(v), v\in V\backslash\{\star\}\}$. A cycle $c_i$ of length $|c_i|=l_i\geq 2$ is called a \emph{proper cycle}. The proper cycles on $\G$ are of even length only, hence, the number of proper cycles $p$ defines the signature of the permutation $\sigma$, that is $\mathrm{sgn}(\sigma)=(-1)^p$. Thus Eq. (\ref{Leibniz}) can be written as follows:
\begin{eqnarray}
\label{Leibniz2}
\det\Delta=\prod_{i=1}^n z_i + \sum_{p=1}^{[n/2]} (-1)^p \sum_{\sigma =c_1 \circ \ldots\circ c_p}\prod_{i=1}^{p}\Delta_{v_i,c_i(v_i)}\Delta_{c_i(v_i),c_i^2(v_i)}\ldots \Delta_{c_i^{l_i-1}(v_i),v_i}\prod_{j \in S_{fp}(\sigma)}z_j
\end{eqnarray}
where $c^k_i$ is the $k$-fold composition of the cyclic permutation $c_i$ of even length $l_i$, $v_i \in \mathcal{O}_i(\sigma)$, so that  $c_i^{k-1}(v_i)\neq c_i^k(v_i)$ and $c_{i}^{l_i}(v_i)=v_i$.
The term $\prod_{i=1}^n z_i$ equals to the number of all spanning digraphs $\G_{dir}(\A)$, having the root $\star$. Each of others terms on the right-hand side of Eq.(\ref{Leibniz2}) having a non-zero set of fixed points $S_{fp}\neq \O$ up to a sign equals to $\prod_{j \in S_{fp}(\sigma)}z_j$, because all non-diagonal elements equal to $-1$. That product represents the number of distinct spanning digraphs which have in common the specified cycles $c_1,\ldots,c_p$, and differ in the oriented edges outgoing from vertices $j \in S_{fp}(\sigma)$. These oriented edges may form cycles on their own which do not enter into the list $c_1,\ldots,c_p$. The proper cycles formed by the oriented bonds incident to fixed points of a given permutation $\sigma=c_1\circ c_2\circ  \ldots \circ c_p$ should enter into enlarged list of cycles $c_1\circ c_2\circ \ldots \circ c_p \circ \ldots \circ c_{p'}$, $p'>p$, corresponding to another permutation $\sigma'$.
The expansion (\ref{Leibniz2}) can be interpreted in form of the \emph{inclusion-exclusion principle} \cite{Priez}. Let $c_1,c_2,\ldots,c_m$ be the list of all possible proper cycles on $\G$. We define $A_i$, $i=1,2,\ldots,m$ as the set of all spanning digraphs on $\G$ containing the particular cycle $c_i$ and $A_0$ is the set of all spanning digraphs $\G_{dir}(\A)$. Let $A_{ST}$ be the set of spanning trees on $\G$. Then we can write of Eq.(\ref{Leibniz2}) in the form:
\begin{eqnarray}
\label{Leibniz3}
\det\Delta=|A_{ST}|=|A_0|-\sum_{i=1}^m |A_i|+\sum_{1\leq i < j \leq m} |A_i\cap A_j|+\ldots+(-1)^m|A_1\cap\ldots\cap A_m|
\end{eqnarray}
where $|A|$ is cardinality of the set A. Eq.(\ref{Leibniz3}) is the Kirhhoff theorem for the number of spanning tree subgraphs of a given graph \cite{Pr85}.

\section{Three-leg correlations}

Now we modify the Laplace matrix changing three non-diagonal elements:
\begin{equation}
\label{DeltaPrime}
\Delta_{ij}' =
\begin{cases}
  z_i & \text{if $i=j$},\\
 -1 & \text{if $i,j$ are nearest neighbors}, \\
-\varepsilon & \text{if $(i,j)\in \mathcal{B}\equiv\{(j_0,t_0),\ (j_0-\hat{x},t_0-\hat{y}),\  (j_0+\hat{y},t_0+\hat{x})$}\} \\
  0 & \text{otherwise}.
\end{cases}
\end{equation}

\begin{figure}[h!]
\includegraphics[width=200pt]{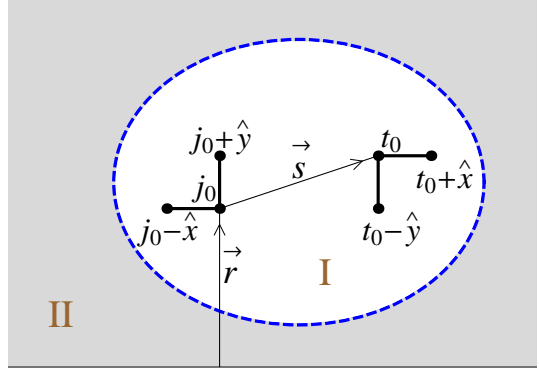}
 \caption{\label{ugolugol2} Component I consists of tree paths and branches attached to them, component II is another spanning tree surrounding component I.}
\end{figure}

In the same determinant expansion as Eq.(\ref{Leibniz2}), only the  terms containing the product $\Delta_{j_0,t_0}\Delta_{j_0-\hat{x},t_0-\hat{y}}\Delta_{j_0+\hat{y},t_0+\hat{x}}=(-\varepsilon)^3$ survive in the limit $\lim_{\varepsilon \rightarrow \infty}  \det\Delta'/ \varepsilon^3$. Permutations $\sigma=c_1\circ c_2\circ \ldots \circ c_p\ $ corresponding to these terms, contain cycles $c_{p_1}\circ \ldots \circ c_{p_k}$ with directed bonds $\mathcal{B}$. Since sites from $\mathcal{B}$ form angles on the lattice, topologically we can draw only one or three cycle(s) containing these bonds. Thus, expression $\lim_{\varepsilon \rightarrow \infty}  \det\Delta'/ \varepsilon^3$ equals to the number of configurations with following features: (i) each configuration is a two component spanning graph on the plane; (ii) one component consists of three paths connecting the vicinity of  site $j_0$ with the vicinity of site $t_0$ and branches of the spanning tree attached to these paths; (iii) another component is the spanning tree having the root $\star$, and surrounding the first component. The quotient of such configurations and all one-component spanning trees is:
\begin{equation}
\label{DetDelta}
\lim_{\varepsilon \ra
\infty}\frac{\det\Delta'}{(-\varepsilon)^3\det\Delta}=-\lim_{\varepsilon
\ra \infty}\frac{\det(I+\delta G )}{\varepsilon^3}\equiv\det(\Lambda)
\end{equation}
where $\delta=\Delta'-\Delta$
and $G=\Delta^{-1}=G\big[(n_1,m_1);(n_2,m_2)\big]$ is the
Green function, which is defined in thermodynamical limit $M,N\rightarrow \infty$ on the plane as:
\begin{eqnarray}
\label{GreenFunction} %
G\big[(n_1,m_1);(n_2,m_2)\big]_{plane}&=&G(n_1-n_2,m_1-m_2)=
\nonumber
\\
&=&G_{0,0}+ \int\!\!\!\!\int_{-\pi}^\pi \frac{d\alpha
d\beta}{8\pi^2} \frac{e^{i
(n_1-n_2)\alpha + i (m_1-m_2)\beta}-1}{2-\cos\alpha- \cos\beta}\ \ ,
\end{eqnarray}
In Appendix we give more details about this function, including its asymptotics on a long distance for arbitrary direction of the vector between $(n_1,m_1)$ and $(n_2,m_2)$. For the case of the upper half-plane, presence of the boundary changes the Green function:
\begin{eqnarray}
\label{GreenFunctionUHP} %
G_{cl}=\Delta^{-1}_{cl}\big[(n_1,m_1);(n_2,m_2)\big]&=&G(n_1-n_2,m_1-m_2)+G(n_1-n_2,m_1+m_2-1)\\
G_{op}=\Delta^{-1}_{op}\big[(n_1,m_1);(n_2,m_2)\big]&=&G(n_1-n_2,m_1-m_2)-G(n_1-n_2,m_1+m_2)
\end{eqnarray}
Matrices in Eq. (\ref{DetDelta}) are of size $n\times n$ (or $n_+ \times n_+$ for the upper half-plane), but, matrix $\delta$ has only three non-zero elements:
\begin{eqnarray}
\nonumber &&\hspace{0.3 in}
\begin{array}{cccc}
 j_0 \ \ \ \  \ & j_0-\hat{x} &j_0+\hat{y} %
\end{array}\\
\label{delta_Ugol}&& \left(
\begin{array}{ccc}
 \ \ \ -\varepsilon\ \ &\  \ \ \  0\ \           &\  \ \   0\ \ \\
 \ \ \   0\ \         &\  \ \ \  -\varepsilon\ \ &\  \ \  0\  \ \\
 \ \ \   0 \ \        &\  \ \ \   0 \ \         &\  \ \ -\varepsilon\ \ \\
\end{array}
\right)
\begin{array}{c}
t_0\\ t_0-\hat{y}\\ t_0+\hat{x}
\end{array}
\end{eqnarray}
so, we obtain due to (\ref{DetDelta}) the matrices of size $3\times 3$ only:
\begin{equation}
\label{Lambda}
\Lambda_{plane}=\left(
 \begin {matrix}
G_{ k , l }     &G_{k+1, l }      &G_{ k ,l-1} \\
G_{ k ,l-1}     &G_{k+1,l-1}      &G_{ k ,l-2}\\
G_{k+1, l }     &G_{k+2, l }      &G_{k+1,l-1}
 \end {matrix} \right).
\end{equation}
\begin{equation}
\label{LambdaClosed} \Lambda_{cl}=\left(
 \begin {matrix}
G_{ k , l }+G_{ k , l+2r-1 }  &G_{k+1, l }+G_{k+1, l+2r-1 }    &G_{ k ,l-1}+G_{ k ,l+2r} \\
G_{ k ,l-1}+G_{ k ,l+2r-2}    &G_{k+1,l-1}+G_{k+1,l+2r-2}      &G_{ k ,l-2}+G_{ k ,l+2r-1}\\
G_{k+1, l }+G_{k+1, l+2r-1 }  &G_{k+2, l }+G_{k+2, l+2r-1 }    &G_{k+1,l-1}+G_{k+1,l+2r}
 \end {matrix} \right)
\end{equation}
\begin{equation}
\label{LambdaClosed} \Lambda_{op}=\left(
 \begin {matrix}
G_{ k , l }-G_{ k , l+2r }  &G_{k+1, l }-G_{k+1, l+2r }   &G_{ k ,l-1}-G_{ k ,l+2r+1} \\
G_{ k ,l-1}-G_{ k ,l+2r-1}    &G_{k+1,l-1}-G_{k+1,l+2r-1}     &G_{ k ,l-2}-G_{ k ,l+2r}\\
G_{k+1, l }-G_{k+1, l+2r }  &G_{k+2, l }-G_{k+2, l+2r }   &G_{k+1,l-1}-G_{k+1,l+2r+1}
 \end {matrix} \right)
\end{equation}
For further analysis we find it convenient to split determinant expansions into four groups:
\begin{eqnarray}
\label{LambdaCl} %
\det\Lambda_{cl}(k,l,r)&=&F^{(3,0)}(k,l)-F^{(2,1)}(k,l,z)+F^{(1,2)}(k,l,z)-F^{(0,3)}(k,z)
\\
\label{LambdaOp} %
\det\Lambda_{op}(k,l,r)&=&F^{(3,0)}(k,l)+F^{(2,1)}(k,l,z)+F^{(1,2)}(k,l,z)+F^{(0,3)}(k,z)
\end{eqnarray}
where $z=l+2r$ for open boundary and $z=l+2r-1$ for closed one.

Functions  $F^{(\mu,\nu)}$ are:
\begin{eqnarray}
\label{Lambdapl} %
 F^{(3,0)}(k,l)
\equiv
 \det\Lambda_{plane}(k,l)
\end{eqnarray}

\begin{eqnarray}
\label{G2klGm} %
F^{(2,1)}&&\hspace{-23pt}(k,l,z)\equiv -G_{k,z}G_{k+1,l-1}^2
+G_{k,z}G_{k,l-2}G_{k+2,l} +G_{k,z-1}G_{k+1,l}G_{k+1,l-1}
-G_{k,z-1}G_{k,l-1}G_{k+2,l} \nonumber \\
&-&2G_{k+1,z}G_{k+1,l}G_{k,l-2}
+2G_{k+1,z}G_{k,l-1}G_{k+1,l-1}-G_{k+1,z-1}G_{k,l}G_{k+1,l-1}+ \nonumber \\ %
&+&G_{k+2,z}G_{k,l}G_{k,l-2}-G_{k+2,z}G_{k,l-1}^2+G_{k+1,z-1}G_{k+1,l}G_{k,l-1}- \nonumber \\ %
&-&G_{k+1,z+1}G_{k,l}G_{k+1,l-1}-G_{k,z}G_{k,l}G_{k+2,l}+G_{k+1,z+1}G_{k+1,l}G_{k,l-1}- \nonumber \\ %
&-&G_{k,z+1}G_{k,l-1}G_{k+2,l}-G_{k,z}G_{k+1,l}^2+G_{k,z+1}G_{k+1,l}G_{k+1,l-1}
\end{eqnarray}
\begin{eqnarray}
\label{GklG2m} %
F^{(1,2)}&&\hspace{-23pt}(k,l,z)\equiv %
-G_{k,z}G_{k+2,z}G_{k,l-2}+G_{k+1,z-1}G_{k,z}G_{k+1,l-1}-G_{k,z-1}G_{k+1,z}G_{k+1,l-1}+ \nonumber \\ %
&+&G_{k,z-1}G_{k+2,z}G_{k,l-1}+G_{k+1,z}^2G_{k,l-2}-G_{k+1,z-1}G_{k+1,z}G_{k,l-1}-G_{k,z}G_{k+2,z}G_{k,l}+ \nonumber \\ %
&+&G_{k,z}G_{k+1,z+1}G_{k+1,l-1}+G_{k,z}^2G_{k+2,l}-G_{k,z-1}G_{k+1,z+1}G_{k+1,l}+2G_{k+1,z}G_{k,z}G_{k+1,l}+ \nonumber \\ %
&+&G_{k,z-1}G_{k,z+1}G_{k+2,l}-G_{k+1,z}G_{k,z+1}G_{k+1,l-1}+G_{k,z+1}G_{k+2,z}G_{k,l-1}+ \nonumber \\ %
&+&G_{k+1,z-1}G_{k+1,z+1}G_{k,l}-G_{k+1,z}G_{k+1,z+1}G_{k,l-1}-G_{k,z}G_{k+1,z-1}G_{k+1,l}
\end{eqnarray}
\begin{eqnarray}
\label{G3m} %
F^{(0,3)}(k,z)&\equiv& %
-G_{k,z}G_{k+1,z-1}G_{k+1,z+1}+G_{k,z}^{2}G_{k+2,z}+G_{k,z-1}G_{k+1,z}G_{k+1,z+1}- \nonumber \\ %
&-&G_{k,z-1}G_{k,z+1}G_{k+2,z}-G_{k+1,z}^{2}G_{k,z}+G_{k+1,z}G_{k,z+1}G_{k+1,z-1}
\end{eqnarray}

\begin{figure}[h!]
\includegraphics[width=300pt]{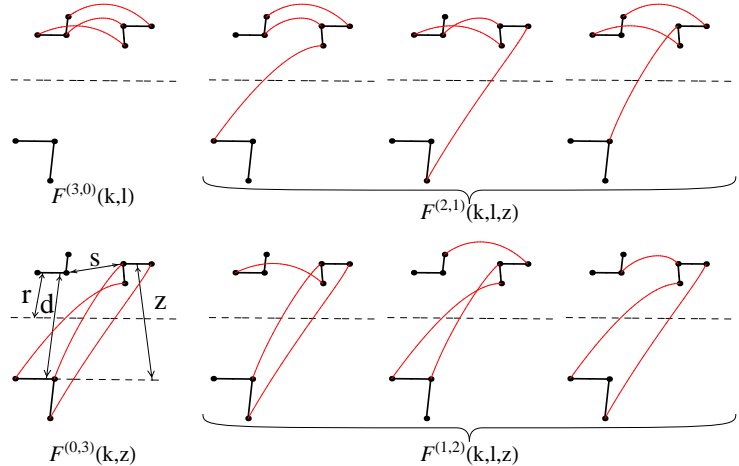}
 \caption{\label{ugolugol3} Angle-angle system near boundary.}
\end{figure}

The meaning of functions $F^{(\mu,\nu)}$ is shown in Fig.\ref{ugolugol3}. The index $\mu=1,2,3$ shows numbers of ``short'' links between the vicinity of  site $j_0$ and the vicinity of  site $t_0$ and index $\nu=1,2,3$ shows number of ``long'' links between the vicinity of the mirror image $j_0'$ of site $j_0$ and the vicinity of $t_0$.

 Using the asymptotic formula for the Green function (Appendix, (\ref{GreenFunctionAsymptAppendix})) we can analyze behavior of functions $F^{(\mu,\nu)}$ for $s\equiv|\vec{s}| \gg1$ and $|\vec{s}+\vec{d}|\gg1$, where $\vec{s}=(k,l)$, $\vec{d}=(0,d)$; $d=2 r$ for the case of UHP with the open boundary and $d=2r-1$ for the closed one:

\begin{eqnarray}
\label{F30Asimpt} %
F^{(3,0)}(k,l)&=&-\frac{1}{8\pi^3}\frac{1+\ln s +c_0 }{s^4}+\ldots \\
\label{F21Asimpt}
F^{(2,1)}(k,l,l+d)&=&-\frac{1}{8\pi^3}\frac{2\cos^2 \beta-1}{s^2 |\vec{s}+\vec{d}|^2}+
\frac{1}{4\pi^3}\frac{\cos \alpha \cos \beta}{s^3 |\vec{s}+\vec{d}|}
+\frac{1}{8\pi^3}\frac{\ln |\vec{s}+\vec{d}| +c_0 }{s^4}+\ldots\ \  \\
\label{F12Asimpt}
F^{(1,2)}(k,l,l+d)&=&-\frac{1}{8\pi^3}\frac{2\cos^2 \alpha-1}{s^2 |\vec{s}+\vec{d}|^2}+
\frac{1}{4\pi^3}\frac{\cos \alpha \cos \beta}{|\vec{s}+\vec{d}|^3 s}
+\frac{1}{8\pi^3}\frac{\ln s +c_0 }{|\vec{s}+\vec{d}|^4}+\ldots
\label{GGGAsympt}\\
\label{F03Asimpt}
F^{(0,3)}(k,l+d)&=&-\frac{1}{8\pi^3}\frac{1+\ln |\vec{s}+\vec{d}| +c_0 }{|\vec{s}+\vec{d}|^4}+
\ldots
\end{eqnarray}
where $\alpha$ and $\beta$ are angles between horizontal axis and vectors $\vec{s}$, $\vec{d}+\vec{s}$ correspondingly. The constant    is $c_0\equiv-2\pi G_{0,0}+\gamma+\frac{3}{2}\ln 2$.

\begin{figure}[h!]
\includegraphics[width=300pt]{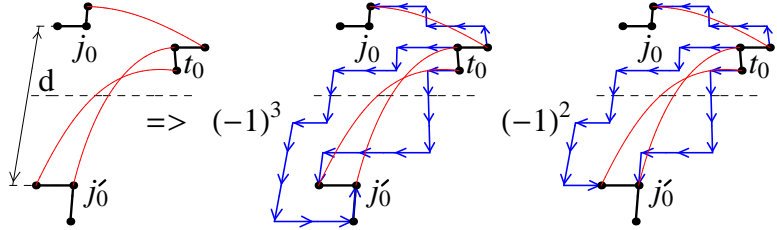}
 \caption{\label{ugolugol4} For large $d\gg s\gg 1$ two configurations are canceled due to asymptotical similarity of paths.}
\end{figure}

The function $F^{(1,2)}(k,l,l+d)$ has the leading term $1/d^2$ for $d\gg s\gg 1$, and its contribution tends to zero as $d\rightarrow \infty$. This function describes configurations with two ``long'' links and one ``short'' link. We can see from Fig.\ref{ugolugol4}, that such configurations define two types of spanning tree subgraphs with opposite sign (this sign changed because the number of loops containing ``long'' links is changed by one). Paths on the square lattice in these two types of subgraphs are topologically almost identical, and differ only in the vicinity of the site $j_0'$. The difference disappears, when $d\rightarrow \infty$.

Now we consider expansions  (\ref{F30Asimpt})-(\ref{F03Asimpt}) of (\ref{LambdaCl}),(\ref{LambdaOp}) in two ways: first we expand these expressions by $r$ assuming
$r\gg s \gg 1$; second we expand them by $s$ assuming $s\gg r \gg 1$. In both cases we take only $\alpha=0$ or $\alpha=\pi/2$ for simplicity. As a result, we get for $r\gg s \gg 1$, $\alpha=0$:
\begin{eqnarray}
\label{AsClosers1Alpha0}
&&\hspace{-5pt}\det\Lambda_{cl}(s,r)=-\Bigg(\frac{1+\ln(2r)+\ln s+2c_0}{8 \pi^3 s^4} +O\left(\frac{1}{s^5}\right) \Bigg)+%
\frac{1}{16\pi^3 r}\Bigg( \frac{1}{s^4}+O\left(\frac{1}{s^5}\right)\Bigg)-
\nonumber \\%
&&\hspace{-26pt}-\frac{3}{8\pi^3 r^2} \Bigg[\frac{3}{8s^2}-\frac{2+\ln(2r)+\ln s+2c_0}{6s^3}+
\frac{\frac{41}{72}+\ln(2r)+\ln s+2c_0}{2s^4}+O\hspace{-3pt}\left(\frac{1}{s^5}\right)\hspace{-2pt}\Bigg]+O\hspace{-3pt}\left(\frac{1}{r^3}\right)
\end{eqnarray}
and for $\alpha=\frac{\pi}{2}$
\begin{eqnarray}
\label{AsClosers1AlphaPi2}
&&\hspace{-15pt}\det\Lambda_{cl}(s,r)=-\Bigg(\frac{1+\ln(2r)+\ln s+2c_0}{8 \pi^3 s^4} +O\left(\frac{1}{s^5}\right) \Bigg)-\frac{1}{16\pi^3 r}\Bigg( \frac{1}{s^3}+\frac{1}{s^4}+O\left(\frac{1}{s^5}\right)\Bigg)+\nonumber \\%
&&\hspace{-20pt}+
\frac{3}{8\pi^3 r^2} \Bigg[\frac{1}{24s^2}-\hspace{-1pt}\frac{1+\ln(2r)+\ln s+2c_0}{6s^3}
-\frac{\frac{53}{72}+\ln(2r)+\ln s+2c_0}{2s^4}+O\hspace{-1pt}\left(\frac{1}{s^5}\right)\hspace{-3pt}\Bigg]\hspace{-1pt}+O\hspace{-1pt}\left(\hspace{-1pt}\frac{1}{r^3}\hspace{-1pt}\right)
\end{eqnarray}
For $s\gg r \gg 1$ and $\alpha=0$ we obtain:
\begin{eqnarray}
\label{AsClosesr1Alpha0}
&&\det\Lambda_{cl}(s,r)=- \frac{3+2c_0+2\ln s}{\pi^3s^{6}}\bigg( r^2-r \bigg)+
\nonumber \\%
&+&\frac{2}{\pi^3s^{8}}\bigg((c_0+\ln s)(6r^4-12r^3+r^2+5r)+\frac{10}{3}(3r^4-6r^3+r^2+2r) \bigg)
+O\left(\frac{1}{s^{9}}\right)
\end{eqnarray}
and for $\alpha=\frac{\pi}{2}$:
\begin{eqnarray}
\label{AsClosesr1AlphaPi2}
 \det\Lambda_{cl}(s,r)&=&-\frac{r}{\pi^3s^{5}}(1+2c_0+2\ln s)+
 \nonumber \\%
&+&\frac{1}{\pi^3s^{6}}\bigg(2(c_0+\ln s)(5r^2-4r)+(3r^2-2r) \bigg)
+O\left(\frac{1}{s^{7}}\right)
\end{eqnarray}
 Asymptotics of $\det\Lambda_{op}(s,r)$ for open boundary conditions for $r\gg s \gg 1$, $\alpha=0$ is
\begin{eqnarray}
\label{AsOpenrs1Alpha0}
&&\det\Lambda_{op}(s,r)=-\Bigg(\frac{1-\ln(2r)+\ln s}{8 \pi^3 s^4} +O\left(\frac{1}{s^5}\right) \Bigg)+%
\nonumber\\%
&+&\frac{1}{16\pi^3 r^2} \Bigg(\frac{5}{4s^2}+\frac{\ln(2r)-\ln s}{s^3}-\frac{\frac{49}{24}+3\ln(2r)-3\ln s}{s^4}+O\left(\frac{1}{s^5}\right)\Bigg)+O\left(\frac{1}{r^3}\right)
\ \ \ \ %
\end{eqnarray}
For $\alpha=\frac{\pi}{2}$ we have
\begin{eqnarray}
\label{AsOpenrs1AlphaPi2}
&&\det\Lambda_{op}(s,r)=-\Bigg(\frac{1-\ln(2r)+\ln s}{8 \pi^3 s^4}+O\left(\frac{1}{s^5}\right)\Bigg) %
+\frac{1}{8\pi^3 r}\Bigg( \frac{1}{2s^3}+\frac{1}{s^4}+O\left(\frac{1}{s^5}\right)\Bigg)+\nonumber\\%
&+&\frac{1}{16\pi^3 r^2} \Bigg( \frac{3}{4s^2}+\frac{\frac{1}{2}-\ln(2r)+\ln s}{s^3}-\frac{\frac{13}{24}+3\ln(2r)-3\ln s}{s^4}+O\left(\frac{1}{s^5}\right)\Bigg)+O\bigg(\frac{1}{r^3}\bigg)
\ \ \ \ %
\end{eqnarray}
In the range of $s\gg r \gg 1$ we have for $\alpha=0$:
\begin{eqnarray}
\label{AsOpensr1Alpha0}
\det\Lambda_{op}(s,r)&=& \frac{2(r^4-5 r^6 + 4r^8)}{3\pi^3s^{12}}+
\frac{74 r^4-394r^6+416 r^8 -96r^{10}}{3\pi^3s^{14}}
+O\left(\frac{1}{s^{15}}\right)
\end{eqnarray}
and for $\alpha=\frac{\pi}{2}$:
\begin{eqnarray}
\label{AsOpenAlphaPi2}
\det\Lambda_{op}(s,r)= \frac{4(r^2+3r^3 + 2r^4)}{3\pi^3s^{8}}+\frac{10 r^2-2r^3-76 r^4 -64r^{5}}{3\pi^3s^{9}}
+O\left(\frac{1}{s^{10}}\right)
\ \ \ %
\end{eqnarray}

\section{Discussion}
\begin{figure}[h!]
\includegraphics[width=450pt]{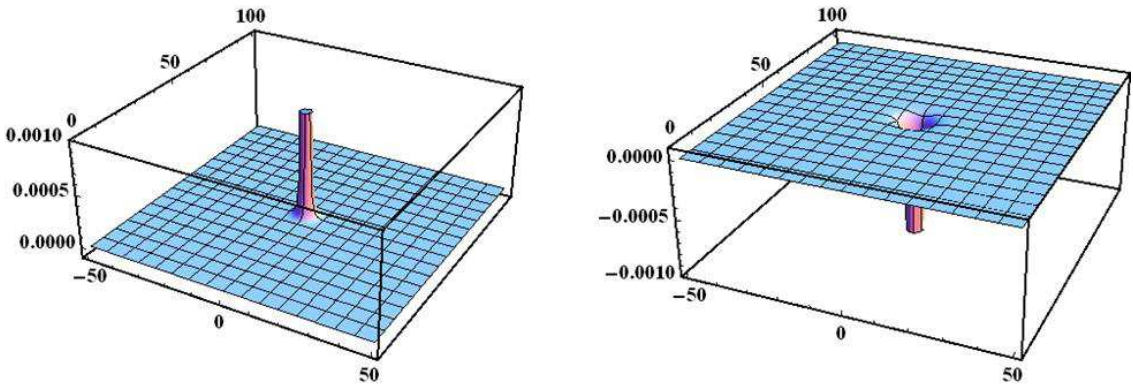}
 \caption{\label{ugolugol5}  $\det\Lambda_{op}(\vec{s},\vec{r})$ and  $\det\Lambda_{cl}(\vec{s},\vec{r})$ for $\vec{s}=(k,l)$, $r=50$, $-50\leq k \leq 50$,  $-50 \leq l \leq 50$.}
\end{figure}

 As it was noticed above, the $\Theta$-graph is the key object in calculations
of heights variables $h_i\geq 2$ in the ASM. The previous analysis \cite{sand,diss,bound,jpr} shows that ASM belongs to a $c=-2$ minimal model of logarithmic conformal field theory. Height variable $h=1$ is associated with a primary field $\phi(z,\overline{z})$ with conformal weights $(1,1)$ and heights $h=2,\ 3,\  4\ $ behave like its logarithmic-partner $\psi(z,\overline{z})$ with scaling dimension $2$. Let $P_h^{UHP}(r)$ be the probability of height $h=1,2,3,4$ at distance $r$ apart from the boundary  of the upper half-plane. In fact this is a two point correlation function, and by mixing operators, its dependence on $r$ enables to obtain a structured constant for the operator product expansion (OPE) and to conjecture the logarithmic behavior of two-point correlations $w_{h_i,h_j}(r)=P_{h_i,h_j}(r)-P_{h_i}P_{h_j}$ of height variables $h_i\geq 1$ on the plane at site $i$ and height variables $h_j\geq 2$ at site $j$ at distance $r$ apart from the site $j$. Correlations $ P_h^{UHP}(r)$ have been obtained combinatorially, by mapping ASM onto the spanning trees model \cite{jpr}.

Due to similarity between the $\Theta$-graph and the three-legs correlations, we may compare
calculations in \cite{jpr} with the present ones. The ``head'' of  $\Theta$-graph located at site
$i$ with $h_i=2$ corresponds to one of the fixed points of the three-legs correlations,
the running point of $\Theta$-graph corresponds to another point at distance $s$ apart.
The dependence of $\det \Lambda(s,r)$ on $s$ for fixed $r$ is shown in
Fig.\ref{ugolugol5} for open and closed boundary conditions. We see
that the function has a strong peak at $(k,l)=(0,0)$. An essential part of computational work in \cite{jpr}
is summation over all positions of the running point. Instead, we may try to use the expansions
(\ref{F30Asimpt})-(\ref{F03Asimpt}) for integration in the vicinity of the peak using the fact that  $\det \Lambda(s,r)$ decays as $\ln s/s^4$ with $s$.

In the case of two-point correlations $w_{1,2}(r)$, this method leads to drastic simplification
of calculations due to rapid convergence of the integrals over the vicinity $s<s_0$ of the peak
for $s_0<<r$  \cite{lette}. In the case of boundary correlations $P_2^{UHP}(r)$, the leading term
of asymptotics by $r$ in (\ref{AsClosers1AlphaPi2}) and (\ref{AsOpenrs1AlphaPi2}) is
$\ln r/r^2$ and its integration by $s$ in the vicinity of the peak is not sufficient for
obtaining a coefficient at $\ln r/r^2$. Indeed, the first terms of expansions
(\ref{AsClosers1AlphaPi2}) and (\ref{AsOpenrs1AlphaPi2}) contain $\ln s/s^4$ which gives
also $\ln r/r^2$ upon summation over the half-plane:
\begin{eqnarray}
\label{intlnss4}
\sum_{k=-\infty}^{\infty}\sum_{l=-r}^{\infty}\frac{\ln |\vec{s}|}{|\vec{s}|^4}\sim \frac{ \ln r}{r^2}.
\end{eqnarray}
 But expansions
(\ref{AsClosers1AlphaPi2}) and (\ref{AsOpenrs1AlphaPi2}) are not valid for $s \sim r$ and
therefore the method  \cite{lette} fails in the case of calculations of $P_2^{UHP}(r)$.

Despite the failure with the description of the  $\Theta$-graph near the boundary, the expansion
(\ref{AsClosers1Alpha0})-(\ref{AsOpenAlphaPi2}) are still useful for a LCFT treatment. The previous attempts to describe the logarithmic correlations both in the lattice and field theories were undertaken solely for the ASM. In this case, the logarithmic partner of the primary field is associated  with the height variables $h_i>1$ having a non-local representation in the spanning tree model. Moreover, the non-local representation is the infinite sum over positions of the running point of the $\Theta$-graph. But two branching points of the $\Theta$-graph are
in turn some correlating objects of the spanning tree which can be considered
in the framework of the LCFT independently of the ASM problem. Thus the collection of asymptotics
(\ref{AsClosers1Alpha0})-(\ref{AsOpenAlphaPi2}) for the three-leg correlations near closed and open boundaries, together with the bulk asymptotics (\ref{F30Asimpt}) should be found within
the LCFT provided that one finds a proper identification for the three-leg branching points.

\section{Appendix}

We consider here the Green function $G_{p,q}$ for two-dimensional square lattice and derive its asymptotic expansion for large distances $\left(r=\sqrt{p^2+q^2}\gg 1\right)$ using the methods proposed in \cite{jpr} and in \cite{Cserti}. We denote an angle between the vector $\vec{r}=(p,q)$ and the horizontal axis  by $\varphi$, so that
\begin{eqnarray}
p \equal r \cos \varphi,\\
q \equal r \sin \varphi.
\end{eqnarray}
Since the Green function contains singular part $G_{0,0}$, we consider a function $g_{p,q} = G_{p,q} - G_{0,0}$,
which has an integral representation
\begin{equation}
g_{p,q}= \frac{1}{8\pi^2}\int\!\!\!\!\int_{-\pi}^{\pi} \; \frac{e^{ {\rm i} p \alpha + {\rm i} q \beta} - 1}{2-\cos\alpha-\cos\beta}
{\rm d} \alpha\; {\rm d} \beta.
\end{equation}
and obeys the symmetry relations,
\begin{equation}
g_{p,q}=g_{-p,q}=g_{p,-q}=g_{q,p},
\end{equation}
so we can put $p > q \geq 0$ without loss of generality.
After the integration over $\alpha$ and symmetrization by $\beta$ we come to the expression
\begin{equation}
g_{p,q} =
\frac{1}{2\pi} \int_{0}^{\pi} \;
\frac{\left(2-\cos \beta-\sqrt{(2 - \cos \beta)^2-1}\right)^{p}\;\cos q \beta\; - 1}{\sqrt{(2 - \cos \beta)^2-1}} \, {\rm d} \beta.
\label{ApInegr1}
\end{equation}
Consider the Taylor expansion of the function
\begin{equation}
e^{p \beta } \left(2-\cos \beta -\sqrt{(2-\cos \beta )^2-1}\right)^p
\end{equation}
for small positive $\beta$ up to order $14$ and denote it $Q(\beta)$:
\begin{eqnarray}
Q_p(\beta)\,  &=& \, 1+\frac{p \beta ^3}{12}-\frac{p \beta ^5}{96}+\frac{p^2 \beta ^6}{288}+\frac{79 p \beta^7}{40320}-\frac{p^2 \beta ^8}{1152}
+\frac{ \left(112 p^3-493p\right) \beta^9}{1161216}+\nonumber\\
&+&\frac{421 p^2 \beta ^{10}}{1935360}
+\frac{ \left(127741p-46200 p^3\right) \beta^{11}}{1277337600}+\frac{ \left(140 p^4-3887p^2\right) \beta ^{12}}{69672960}+\nonumber\\
&+&\frac{ \left(69432p^3-152461p\right) \beta ^{13}}{6131220480}
+\frac{\left(629861 p^2 -43120 p^4\right) \beta^{14}}{42918543360} \, .
\end{eqnarray}
The function
\begin{equation}
R_p(\beta) =\left| \left(2-\cos \beta -\sqrt{(2-\cos \beta )^2-1}\right)^p - e^{- p \beta }Q(\beta) \right|
\end{equation}
vanishes at $\beta=0$, and has a unique maximum over $\beta$ for all $p$.
The location of the maximum, $\beta^*$, can be found as a series in $\frac{1}{p}$
if we construct the series expansion of the derivative $\frac{\d R_p(\beta)}{\d\beta}$ and recursively equate coefficients to 0.
Calculations give
\begin{equation}
\beta^* = \frac{15}{p} + \frac{795}{28 p^3} + \frac{14503977}{17248 p^5} + \cdots
\end{equation}
and
\begin{equation}
R_p(\beta) \leq \frac{120135498046875}{8192\, e^{15}\, p^{10}} + O\left(p^{-12}\right)
\end{equation}
It is easy to show that
\begin{equation}
\int_{0}^{\pi} \beta ^n e^{-p \beta } \cos (q\beta) \frac{{\rm d} \beta}{2 \pi } \sim \frac{1}{r^{n+1}}, \quad n \geq 0
\end{equation}
It means that we can replace the function in the integral (\ref{ApInegr1}) by its Taylor expansion in $\beta$ up to $14$-th
order and get an expression for the Green function with accuracy $O(p^{-10})$.
It is convenient to express the Green function as a sum of three integrals
\begin{equation}
g_{p,q}=I_1+I_2+I_3,
\end{equation}
where
\begin{equation}
I_1=\frac{1}{2 \pi }\int_0^{\pi } \left(\frac{e^{-p \beta } \cos (q \beta )}{\beta }-\frac{1}{\beta }\right) \,
   {\rm d} \beta
\end{equation}

\begin{equation}
I_2=\frac{1}{2\pi}\int_0^{\pi } \left(\frac{1}{\beta }-\frac{1}{\sqrt{(2-\cos \beta)^2-1}}\right) \, {\rm d} \beta
\end{equation}

\begin{equation}
I_3=\int_{0}^{\pi} \;\left(
\frac{e^{-p \beta} Q_p(\beta) \;\cos (q \beta)}{\sqrt{(2 - \cos \beta)^2-1}} -\frac{e^{-p \beta } \cos (q \beta )}{\beta }
\right) \frac{{\rm d} \beta}{2\pi} + O\left( \frac{1}{r^{10}} \right).
\label{I3}
\end{equation}
The expressions give
\begin{equation}
I_1 = -\frac{1}{2\pi} (\ln r + \gamma + \ln \pi ) + \text{exponentially small terms}
\end{equation}

\begin{equation}
I_2 = \frac{1}{2 \pi } \left( \ln \pi - \frac{3 \ln 2}{2} \right)
\end{equation}
Finally we obtain
\begin{eqnarray}
\label{GreenFunctionAsymptAppendix}
g_{p,q}\,  &=& \,
-\frac{\ln r + \gamma +\frac{3 \ln 2}{2}}{2 \pi }
+\frac{\cos (4 \varphi )}{24 \pi  r^2}
+\frac{1}{r^4}\left(\frac{3 \cos (4 \varphi )}{80 \pi }+ \frac{5 \cos (8 \varphi )}{96\pi}\right)+\nonumber \\
&+&\frac{1}{r^6}\left(\frac{51 \cos (8 \varphi )}{224 \pi }+\frac{35 \cos (12\varphi )}{144 \pi }\right)
+\nonumber \\
&+&\frac{1}{r^8}\left(\frac{217 \cos (8 \varphi )}{640 \pi }+\frac{45 \cos (12 \varphi )}{16 \pi }+\frac{1925 \cos(16 \varphi )}{768 \pi }\right)
+ O\left( \frac{1}{p^{10}} \right)
\end{eqnarray}
where the expansion
\begin{eqnarray}
\frac{1}{\sqrt{(2-\cos \beta)^2-1}} &=& \frac{1}{\beta }-\frac{\beta }{12}+\frac{43 \beta ^3}{1440}
-\frac{949 \beta ^5}{120960}+\frac{21727\beta ^7}{9676800}-\frac{501451 \beta ^9}{766402560}+ \nonumber \\
&+&\frac{8112267073 \beta^{11}}{41845579776000}-\frac{277899049 \beta ^{13}}{4782351974400}+O\left(\beta ^{15}\right)
\end{eqnarray}
is used for the calculation of (\ref{I3}).

%

\section*{Acknowledgments}
This work was supported by a Russian RFBR grant No 09-01-00271. V.S.P. would like to thank Dynasty foundation for financial support. S.Y.G. is grateful for JINR grant-2009 for young scientists.

\end{document}